\documentclass{article}
\usepackage[dvips]{epsfig}
\usepackage{typearea,cite,overcite,amsmath,amssymb}
\begin{document}
\makeatletter
\def\@cite#1{\mbox{$\m@th^{\hbox{\@ove@rcfont[#1]}}$}}
\makeatother
\title{Chain stiffness intensifies the reptation characteristics of polymer
  dynamics in the melt}
\author{Roland Faller\footnotemark \and Florian M{\"u}ller-Plathe\\
  \small Max-Planck Institut f{\"u}r Polymerforschung, Ackermannweg 10, 55128
  Mainz, Germany}
\maketitle
\setcounter{footnote}{1}
\renewcommand{\thefootnote}{\fnsymbol{footnote}}
\footnotetext{present address: Department of Chemical Engineering, University 
of Wisconsin, Madison, WI 53706, USA}
\begin{abstract}
  The reptation concept in polymer dynamics is studied for model chains with
  added stiffness. The main idea of a chain diffusing inside a tube can be
  transferred from fully flexible chains although the renormalization onto a
  flexible chain of fewer Kuhn segments fails. The entanglement length shrinks
  with increasing persistence length. If entanglement length and persistence
  length come to the same order of magnitude the picture of a tight tube is
  better suited, in which chain segments can move only along the contour, any
  transverse motion being much reduced. Thus, as stiffness increases, the
  monomers loose their freedom to perform random walks inside the tube, the
  ``Rouse-like'' part of their dynamics.
\end{abstract}


\noindent
For the dynamics of polymer chains longer than the characteristic entanglement
length in the melt the concept of reptation~\cite{degennes71,doi86} is widely
accepted. In contrast to short chains which move isotropically and which can
be described by the Rouse model~\cite{rouse53}, longer chains are constrained
to an effective tube by topological entanglements due to the surrounding
chains. The many chain problem may be replaced by an effective single chain
problem with a mean field representation of the neighbors. This model has been
developed for fully flexible polymers and is based on the Rouse model for
local dynamics, which treats a polymer chain as a set of non-interacting beads
connected by springs in a bath of homogeneous friction. The entanglement
monomer number $N_e$ is the chain length at which the cross-over from Rouse to
reptation dynamics occurs.\cite{kremer90} As chain stiffness is
added to the model, the Rouse description is no more appropriate for the local
dynamics~\cite{harnau97}. So it becomes questionable if the standard reptation
picture is still suitable for the dynamics of entangled stiff chains. After
some earlier work on the reptation of {\it almost rigid} rods~\cite{odijk83},
recent theoretical work tries to adapt the reptation model to moderately stiff
polymer chains~\cite{morse98a}.

The effective stiffness of a polymer is characterized by its persistence
length $l_p$ which is defined as the exponential decay length of orientation
correlations. This is the 
chain length for which a polymer keeps its orientation.  Several theoretical
models propose different scalings of the entanglement length with the
persistence length
stiffness~\cite{odijk83,doi85,wu89,sato91a,fetters94,morse98a}. There is not
even consensus whether the entanglement length rises or falls with
stiffness. Scaling behaviors ranging from $N_e\propto l_p^{-3}$ to $N_e\propto
l_p^2$ have been proposed.

The aim of this simulation study is to investigate the qualitative change of
polymer dynamics from the standard Rouse/reptation picture as chain stiffness
is introduced and increased. We show that quantities like the entanglement
length or the tube diameter become somewhat diffuse and ambiguous in this
case. Still, the concept of an entangled polymer moving in a tube holds; and
irrespective of the way of defining the entanglement length, it can be shown
to {\it decrease} with increasing persistence length $l_p$. Eventually, the
two lengths become comparable and the chain dynamics changes to pure reptation
dynamics without an underlying local short-time Rouse regime. One should note
that most real polymers have some chain stiffness and they do show local
dynamics very similar to our simulations as has been shown in recent NMR
experiments~\cite{graf98}. The interplay of entanglements and stiffness is
sometimes discussed in terms of a packing
length~\cite{fetters99a,fetters99b}. The experimentally accessible region is,
however, much smaller than the parameter area discussed in the following,
i.e. in terms of stiffness experimental polymers are less different than they
appear on first sight. Yet, simulations and experiments have shown that the
Rouse model is not fully appropriate for real polymers especially at short
ranges where local properties play an increasing
role~\cite{richter94,paul97,harmandaris98}.

We apply a well-described simulation model of polymer
dynamics\cite{grest86,kremer90,puetz00} augmented with intrinsic
stiffness\cite{micka97,faller99b}
along the polymer backbone. The chains consist of purely repulsive soft
spheres connected by anharmonic springs and a harmonic bond angle
potential. Constant temperature is maintained by Brownian dynamics. For
details of the simulated systems see ref.~\citen{faller00b}. We have
investigated systems containing up to 2000 chains of 2 to 1000 monomers of
different stiffness ranging from fully flexible to a persistence length of
five monomer diameters. The chains with no imposed stiffness (the original
model~\cite{grest86}) have a persistence length of about one monomer diameter
resulting from the excluded volume interaction. The units used in this article
are Lennard-Jones reduced units with monomer mass $m$, monomer diameter
$\sigma$ and interaction strength $\epsilon$ set to 1. For the exact
definition and the potentials see ref.~\citen{faller99b}. As melts with
persistence lengths much larger than five turn nematic~\cite{kolb99}, they
were not investigated. The time-step was set to 0.01 time units, the density
to $\rho=0.85\sigma^{-3}$ and the average bond length turns out to be
$0.97\sigma$.

Mean-squared displacements (MSD) of central monomers in a chain, in the
following referred to as $g_1(t)$, are a key property both in theory and
simulation, as they illustrate the different dynamic regimes of reptation. The
dependence of $g_1$ on time scales according to theory with four dynamical
exponents in the reptation case~\cite{doi86}: at small times the monomers
behave as in the Rouse model ($g_1\propto t^{1/2}$); after the so-called
entanglement time $\tau_{e}$, the motion is constrained to a tube, thus the
Rouse relaxation is one-dimensional along a random walk leading by
transformation to three dimensions to $g_1\propto t^{1/4}$; at the Rouse time
$\tau_R$ the internal degrees of freedom are relaxed but the motion is still
confined to the tube ($g_1\propto t^{1/2}$); and after the chain leaves the
tube at the disengagement time $\tau_d$ free diffusion ($g_1(t)\propto t$) is
observed. This behavior was confirmed by simulations of fully flexible
chains~\cite{kremer90,duenweg98,puetz00}. The first three regimes can be seen
for the $l_p=1$ curve in Figure~1a.

Depending on stiffness we find two different classes of dynamics. Up to a
persistence length of $l_p\le3$, the chains behave qualitatively like
flexible chains. Chains of length $N=200$ are at the density used clearly
entangled for $l_{p}=1.5$ and $l_{p}=3$. Even the chains of length 75 and
persistence length of only 1.5 exhibit a slowdown to $g_1(t)\propto t^{0.3}$
(Figure~1a), which is more pronounced than for fully flexible chains
at $N=200$~\cite{duenweg98}.
\begin{figure}
  \includegraphics[angle=-90,width=0.49\linewidth]{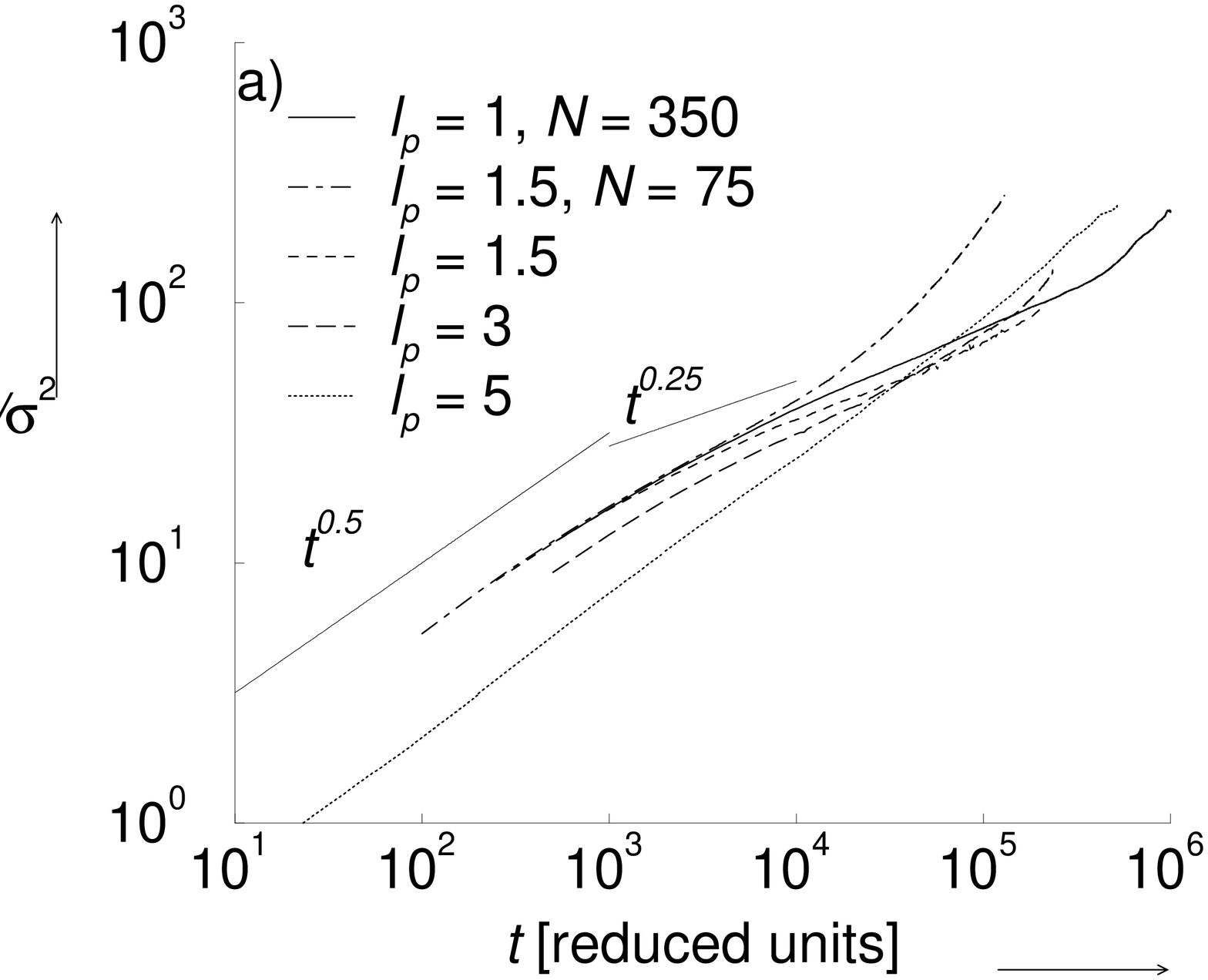}
  \includegraphics[angle=-90,width=0.49\linewidth]{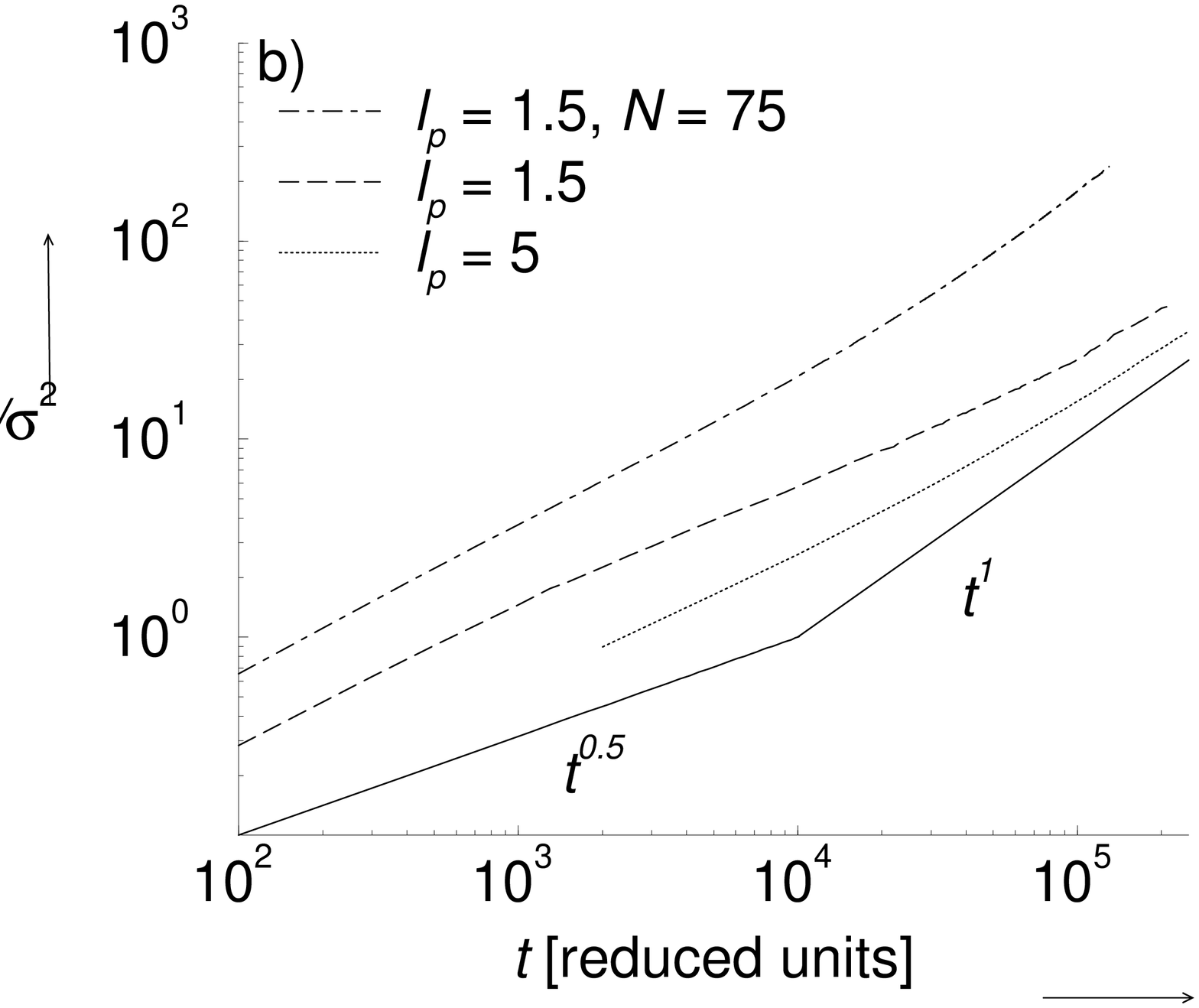}
  \caption{Mean squared displacements of a) monomers in the central part of
    the chain b) center of mass. Chains of length 200, unless stated
    otherwise.}
  \label{fig:msd}
\end{figure}
With increasing persistence length the value of the mean-squared-displacement
value at the crossover $\tau_e$ from $t^{1/2}$ to $t^{1/4}$, which is
associated with the square of the tube diameter $d_T$ shrinks slightly. This
is a first indication that the entanglement length ($\approx d_T^2/l_p$ for
flexible chains) decreases with increasing persistence length. The time
$\tau_e$ can be transformed into the relaxation time of a chain, from which the
entanglement length (i.e. the number of monomers) $N_e$ can be
obtained.\cite{doi86,kremer90} The
Rouse model is standardly employed in this mapping procedure. As - in the
presence of stiffness - the Rouse dynamics is no longer appropriate even
for short chains, $N_e$ becomes somewhat ambiguous.

For the system with persistence length $l_{p}=5$ not even the standard
reptation scenario is recovered, as the $t^{1/4}$ regime is not observed. In
the flexible case the tube diameter $d_{T}$ estimated from the crossover in the
MSD to $t^{1/4}$ is about $\sqrt{40}\sigma\approx 6\sigma$, and in
Figure~1a it can be seen that it decreases further with
$l_p$. Hence, it can be safely assumed that, for the $l_p=5$ chains, the tube
diameter is equal to or smaller than the persistence length. As a result,
stiff chains cannot wiggle freely in their tube but have to follow closely the
primitive path. This is in marked contrast to the original Rouse/reptation
model, in which chains perform a random walk inside their tubes. Now, the
chain can only perform reptation moves because there is no transverse freedom
to explore. On local scales, the stiffness hinders the transverse motion and,
on longer scales, the entanglements prevent it. The tube itself appears
worm-like rather than like a random walk~\cite{morse98a}. It does not differ
much from the chain whose worm-like nature is e.g. visible in its static
structure factors~\cite{faller99b}. Still, the limit of persistence length
large compared to entanglement length is not approached.

Note that the different MSDs intersect in Figure~1a.  The stiffer
chains are {\it faster} in an intermediate regime. For long times (not shown),
however, we expect that the size of the ellipsoid spanned by the chain defines
the ultimate diffusion speed. Thus, the flexible chains should be faster
again.  This is indicated e.g. by comparing the chains with $l_p=3$ and
$l_p=5$, where a second intersection is very probable for longer times
(Figure~1a). A simple argument assuming strictly reptating chains
compared to the standard reptation supports this picture: The final diffusion
depends only on the extension, so that the stiffer chains are slower. However,
for these chains the first two regimes are missing due to the lack of the Rouse
regime. This was confirmed by an analysis of the Rouse 
modes~\cite{faller00a}. They start directly with the third regime so that the
slowdown to $t^{1/4}$ for the flexible chains lets the stiffer chains
overtake~\cite{faller00a}.

The centers of mass of chains can diffuse ``freely'' in the unentangled case
$(N<N_e)$, whereas in the entangled case a sub-diffusive motion arising from
the constraining to the curvilinear tube is found up to the Rouse time
$\tau_R$ (Figure~1b).  The chains with $l_p=5$ and $N=200$ are
highly entangled, which is not only visible by the sub-diffusive nature of
the MSD of the center of mass $g_3(t)$ but also by their reorientation
behavior~\cite{faller00b,faller00a}.

According to the Rouse model, the center of mass diffusion coefficient $D$ for
$t\to\infty$ scales with $N^{-1}$. Deviations from this in the long chain
regime are normally attributed to entanglements~\cite{kremer90}.
Figure~2 shows the dependence of $DN$  on the chain length
for different stiffnesses.
\begin{figure}[t]
  \includegraphics[angle=-90,width=0.49\linewidth]{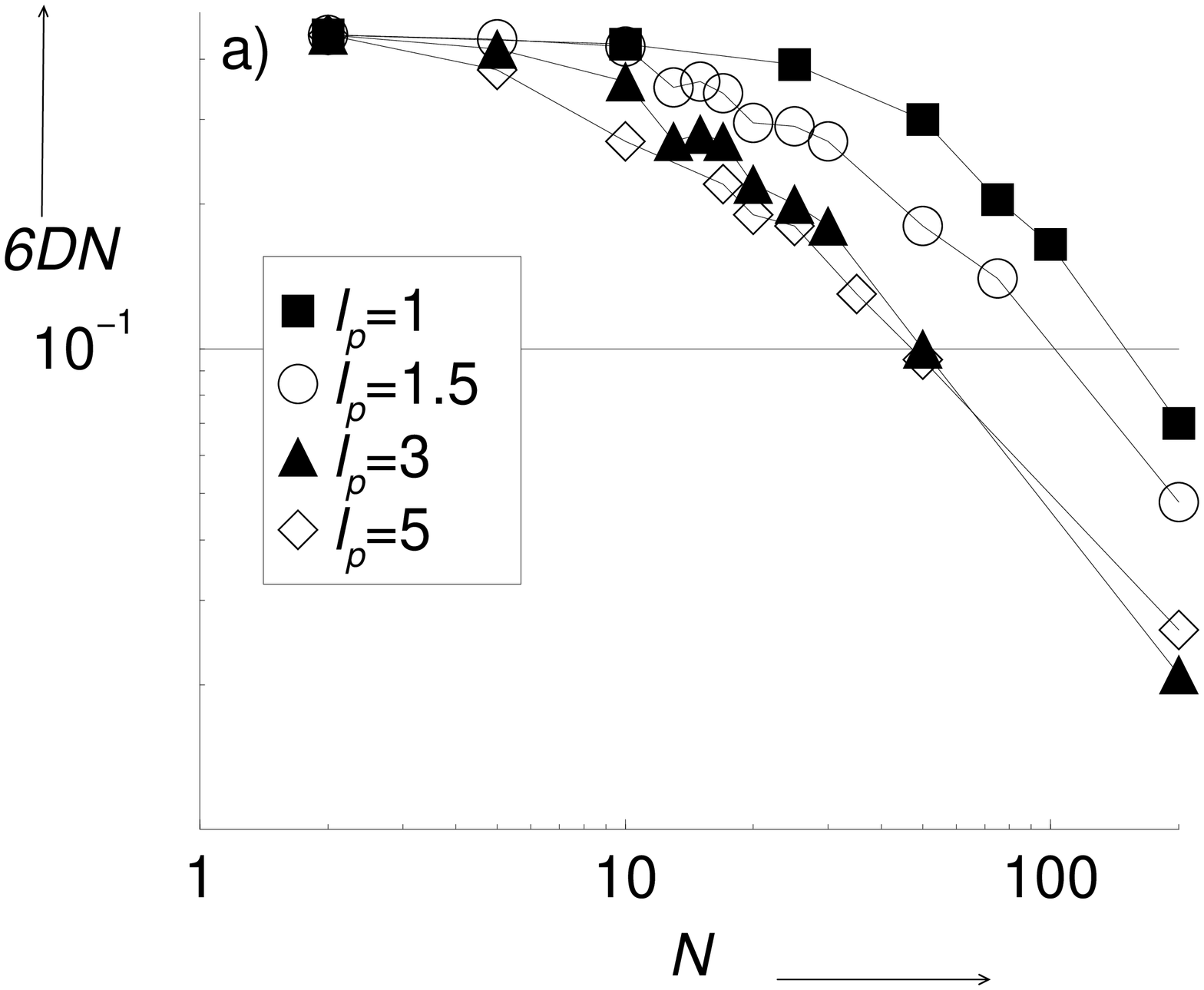}
  \includegraphics[angle=-90,width=0.49\linewidth]{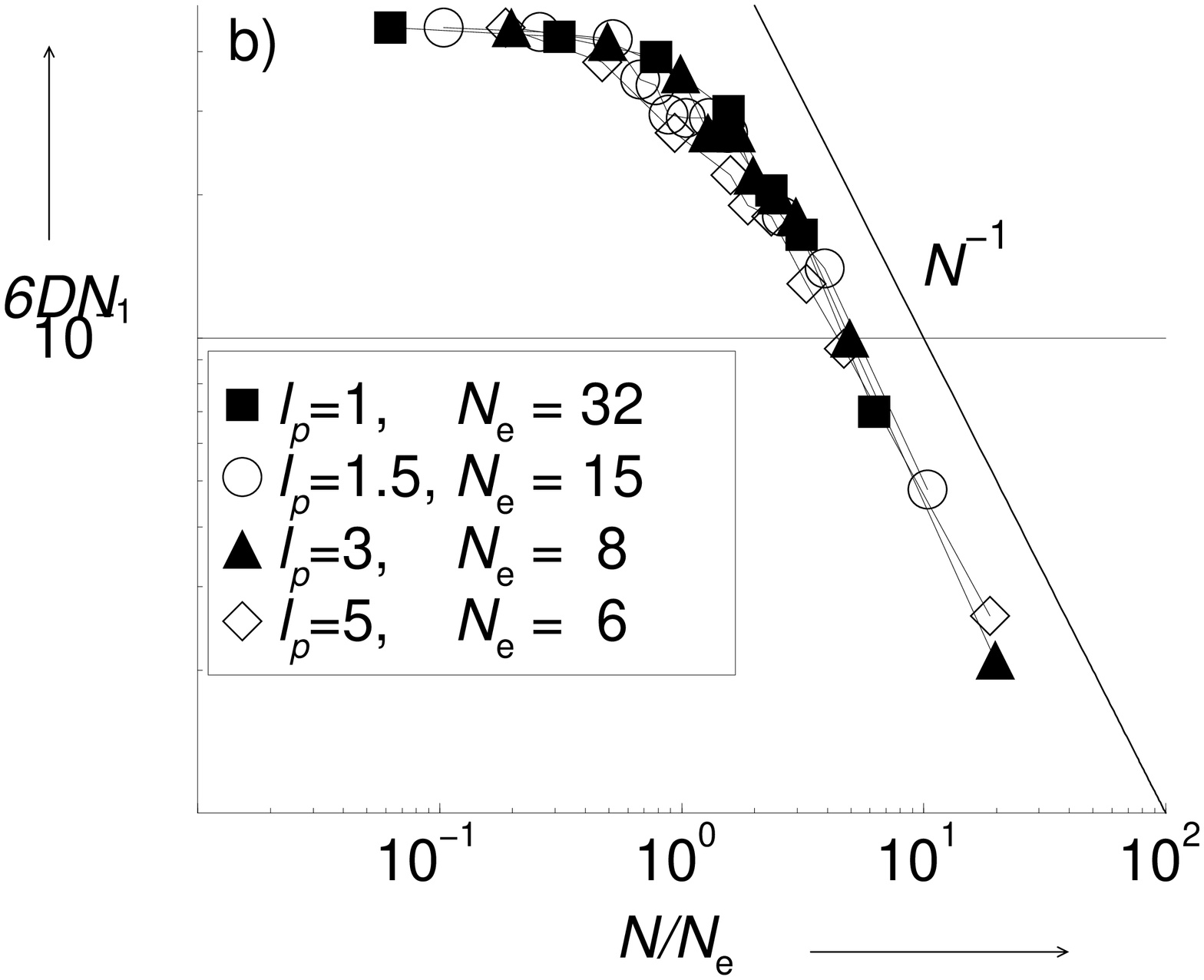}
  \caption{
    a) Center of mass diffusion constant for different chain lengths in
    Rouse scaling, b) same as a) but rescaled empirically to define a
    entanglement length. Below $N=50$ the errors are smaller than the symbol
    sizes. 
    In both figures the filled squares correspond to $l_p=1$, the open circle
    to $l_p=1.5$, the filled triangle to $l_p=3$ and the open diamond to 
    $l_p=5$.
    The entanglement lengths depending on persistence length used in 
    part b) are $N_e(1)=32$, $N_e(1.5)=15$, $N_e(3)=8$, and $N_e(5)=6$. 
    }
  \label{fig:DN}
\end{figure}
The fully flexible model $l_p=1$ shows Rouse behavior up to $N\approx 32$ and
then the diffusion coefficient falls off with $D\propto N^{-2}$. This picture
does not directly extend to higher persistence lengths for which there are
again two classes of dynamics. For persistence length $1.5\le l_{p}\le3$ there
is a first plateau (Rouse) at very short chain lengths ($N\lesssim10$). Then
the diffusion constant decreases faster than $N^{-1}$. A second short plateau
is found ($10\lesssim N\lesssim20$) followed by a decay as the entanglement
length is approached. For higher stiffness ($l_p=5$) no plateau at all is
found, indicating that the Rouse model cannot be applied even for the
shortest chains. For this system, a Rouse scaling of the autocorrelation
functions of the Rouse modes is not possible either~\cite{faller00a}.  The
Rouse regime is ``squeezed out'' between the short local modes and the long
modes hindered by entanglement. With increasing $l_p$ longer wavelengths are
influenced by stiffness. As the entanglement length decreases at the same time
shorter wavelengths become entangled. This eventually annihilates the regime
of random (Rouse) motion.

By rescaling the abscissa to match the diffusion coefficient at $N_e=32$
(Figure~2b) one can derive estimates of entanglement lengths:
$N_e(l_p=1.4)=15$, $N_e(3)=8$, and $N_e(5)=6$. For flexible chains the
entanglement length coincides with the value derived by the crossover in the
$g_1$ function~\cite{kremer90,puetz00}.

The effective entanglement lengths are not as sharply defined as in the
flexible case. Even for fully flexible chains there are several definitions of
the entanglement length~\cite{puetz00}, and this ambiguity becomes worse for
moderately stiff chains which have two plateaus in $DN(N)$. However, the
decrease of $N_e$ with $l_p$ observed is systematic,
independent of the details how $N_e$ is actually extracted. Moreover, the
scaling described above leads to a collapse onto a master curve to a
reasonable approximation (Figure~2b).

The previous section showed that stiffness in connection with topological
entanglements leads to a novel dynamics of chains in the melt. The character
of the local motions can be investigated in more detail. The correlation
function
\begin{eqnarray*}
  \Pi_d(t) = \Big\langle P_{2}\left(\frac{\vec{x}_{i}(t)-\vec{x}_{i}(0)}
  {|\vec{x}_{i}(t)-\vec{x}_{i}(0)|}\cdot
  \frac{\vec{x}_{i+d}(\frac{t}{2})-\vec{x}_{i-d}(\frac{t}{2})}
  {|\vec{x}_{i+d}(\frac{t}{2})-\vec{x}_{i-d}(\frac{t}{2})|}\right)\Big\rangle
\nonumber
\end{eqnarray*}
connects the direction of the displacement of a monomer with the local chain
direction, where $P_2$ is the second Legendre polynomial. The length of the
corresponding chain segment is equal to $2d$. For $d=1$, it is defined by the
two nearest neighbors. This function indicates if the beads move predominantly
in the chain direction or if their motion is not correlated to this
direction. As the highly entangled $l_p=5$
systems have fewer transversal degrees of freedom, the correlation is
high, signaling a dominance of reptation (Figure~3a).
\begin{figure}
  \includegraphics[angle=-90,width=0.49\linewidth]{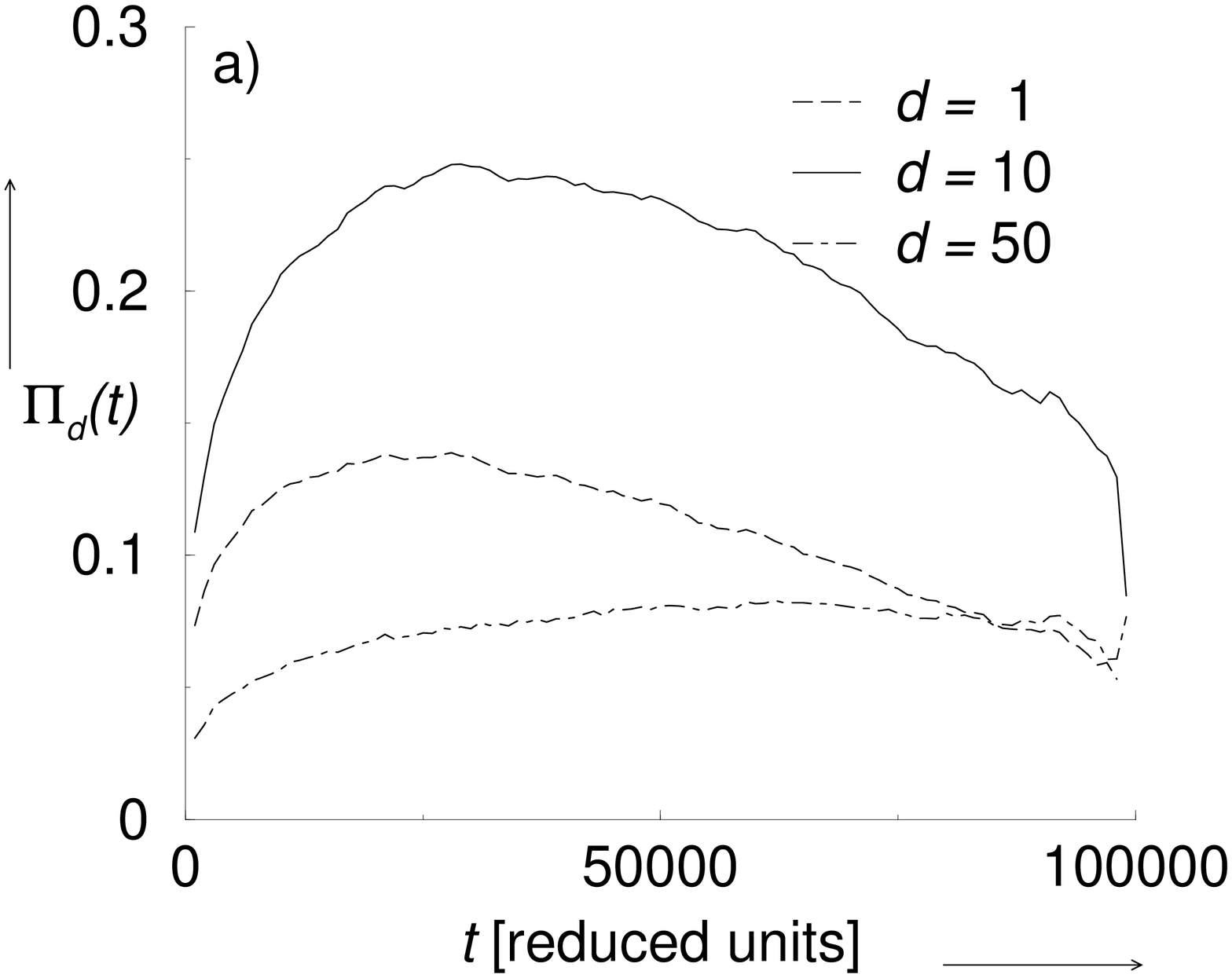}
  \includegraphics[angle=-90,width=0.49\linewidth]{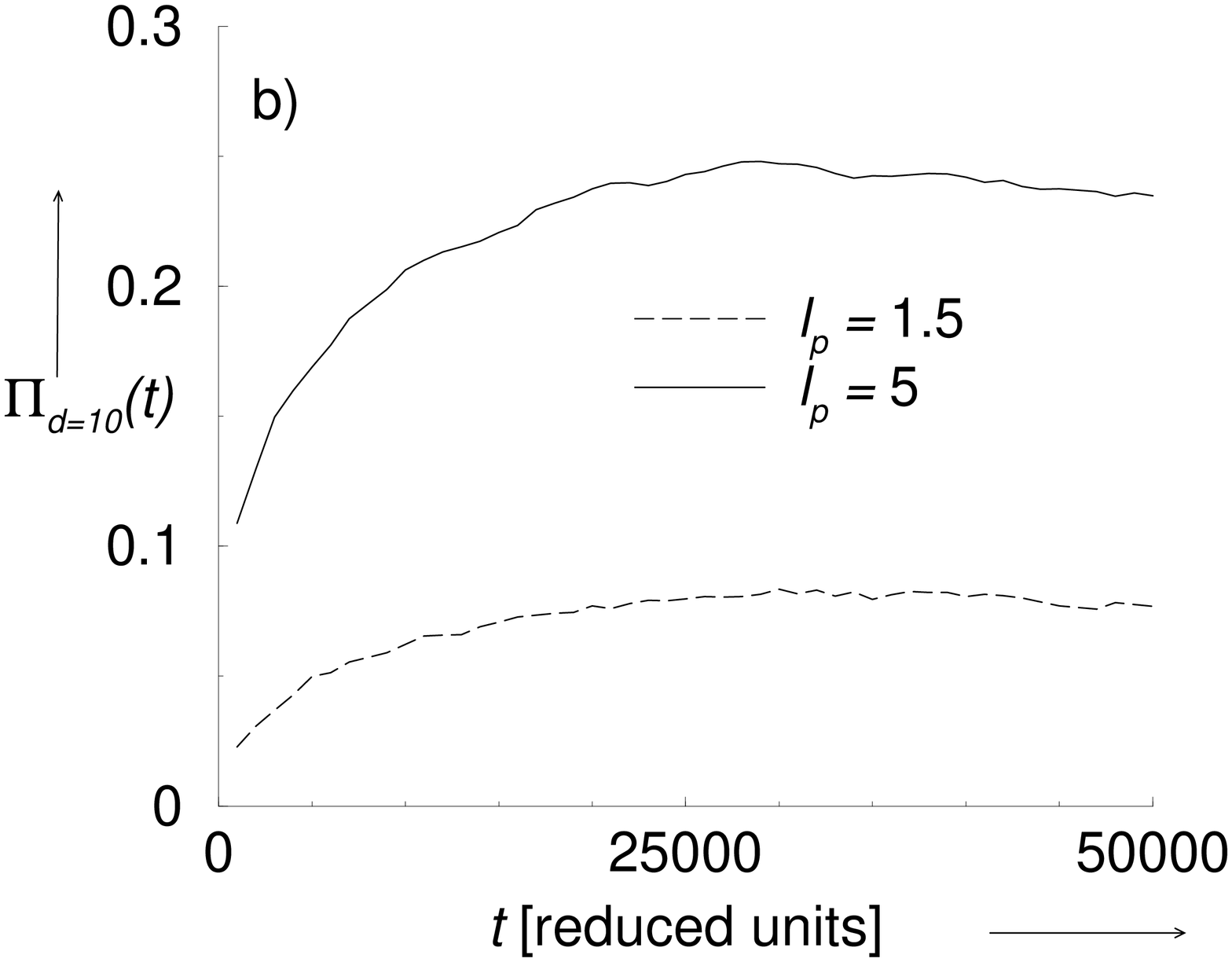}
  \caption{Correlation of inner chain direction vectors with their
    displacements $\Pi_d$ a) 
    Different lengths of segments for $l_p=5,\,N=200$. b) Comparison of
    different stiffnesses $N=200,\,d=10$.}
  \label{fig:corr-conn-disp}
\end{figure}
There is evidence for a finite correlation length as the maximum correlation
is not encountered for $d=1$ at $l_p=5$ (Figure~3a).  For
$d=10$ ($\approx$ Kuhn segment length) the correlation is larger than for
$d=1$ and for $d=50$. At local scales some transverse degrees of freedom in
the tube remain and on the really large scale the {\it static} structure
shines through. As the statics is not influenced by entanglements, it can be
described by a random walk of blobs~\cite{faller99b}.
Figure~3b provides yet more evidence for the
enhancement of reptation by added stiffness: At all times the correlation of
motion with chain direction is stronger for $l_p=5$ than for $l_p=1.5$.

The tube is visualized in Figure~4. The numbers denote the
positions of the chain end in the order they appear. The end clearly drags
back into the tube and moves out again into a new tube. Note that the figure
includes {\it all} monomers without any preaveraging. One can distinguish 
stretched segments of the
tube in which the chain is also stretched and shows almost no crumpling over
the entire simulation. These rigid segments can undergo a subdiffusive
translation while keeping their orientation. Simultaneously reptation in the
tube is possible. This illustrates the concept of ``chain reptation without
Rouse motion'' for stiff polymers.
\begin{figure}
  \[
  \includegraphics[width=0.6\linewidth]{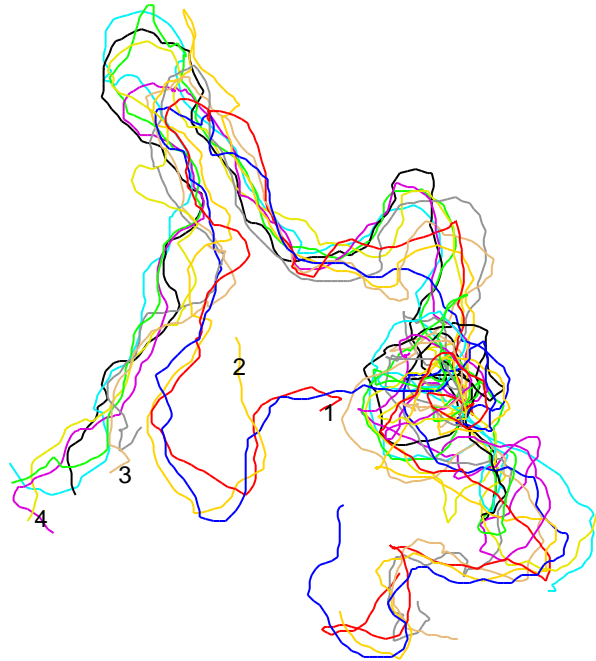}
  \]
  \caption{One chain in the system with $l_p=5$ and $N=200$. The different
    frames are at a time distance of $\Delta t=10000$. 
    }
  \label{fig:rept}
\end{figure}
This expresses itself also in the fact that reorientation correlation
functions of bond-vectors decay algebraically on short time scales. And
monomers propagate through space without losing all of their
correlation~\cite{faller00b}. 

From the dynamic structure factor the tube diameter can be inferred. For Rouse
dynamics, the structure factor decays uniformly, whereas reptation leads to
the development of a plateau. Recent experiments showed the presence of
reptation e.g. in polyethylene melts~\cite{schleger98}.

The dynamic structure factor of the flexible chains can be well fitted by the
formula of Kremer and Binder~\cite{kremer84}.
\begin{eqnarray}
  \frac{S(k,t)}{S(k,0)}&=&\left\{1-\left(\frac{kd_T}{6}\right)^2
    f\left[k^2l_{b}^2\left(\frac{3k_{B}T}
      {l_{b}^2\zeta}t\right)^{1/2}\right]\right\}\times\nonumber\\
  &&\frac{8}{\pi^2}\sum\limits_{p=1}^{\infty}\frac{1}{(2p-1)^2}
  \exp\left(-\frac{t(2p-1)^2}{\tau_d}\right)\;,
  \label{eq:dynstru}\\\nonumber
  f(u)&=&\exp\left(\frac{u^2}{36}\right)\text{erfc}
  \left(\frac{u}{6}\right)\;.
\end{eqnarray}
The only fit parameter employed is the tube diameter $d_T$, the Rouse friction
$\xi$ was obtained from the decay of the Rouse modes, the disengagement time
$\tau_d$ can only be estimated assuming the standard picture. Still, the fit
works reasonably well for $l_p=1.5$ (Figure~5a), resulting in a
tube diameter of 12.2 for chains of length 200 which is most probably too high
compared to the value obtained from the monomer displacements (see above). For
flexible chains it was shown that this discrepancy vanishes only in the limit
of very large $N$~\cite{puetz00}. For the even shorter chains of length 75, we
derive a value of 15.4, which indicates a tube diameter decreasing with
length. It also decreases with stiffness; for $l_p=5$ 
Equation(~\ref{eq:dynstru}) yields 6.1. However, the fit formula is not really 
appropriate to the stiff chains as the local Rouse dynamics on which it is 
based is absent (Figure~5b). Thus, this shows again that these chains are 
subject to a dynamics which cannot be described in terms of standard reptation
theory. Still, we used the formula to yield at least a (rather crude) estimate
of the tube diameter.
\begin{figure}
  \includegraphics[width=0.49\linewidth]{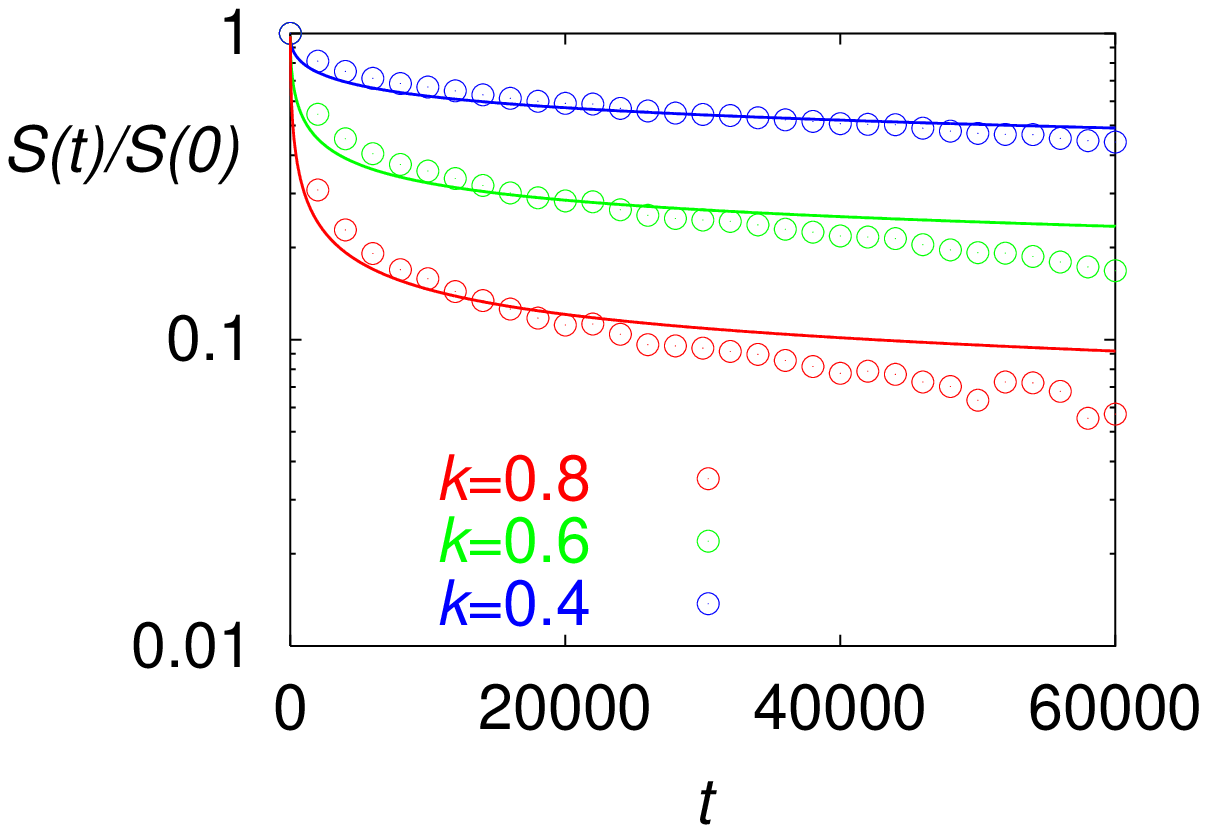}
  \includegraphics[width=0.49\linewidth]{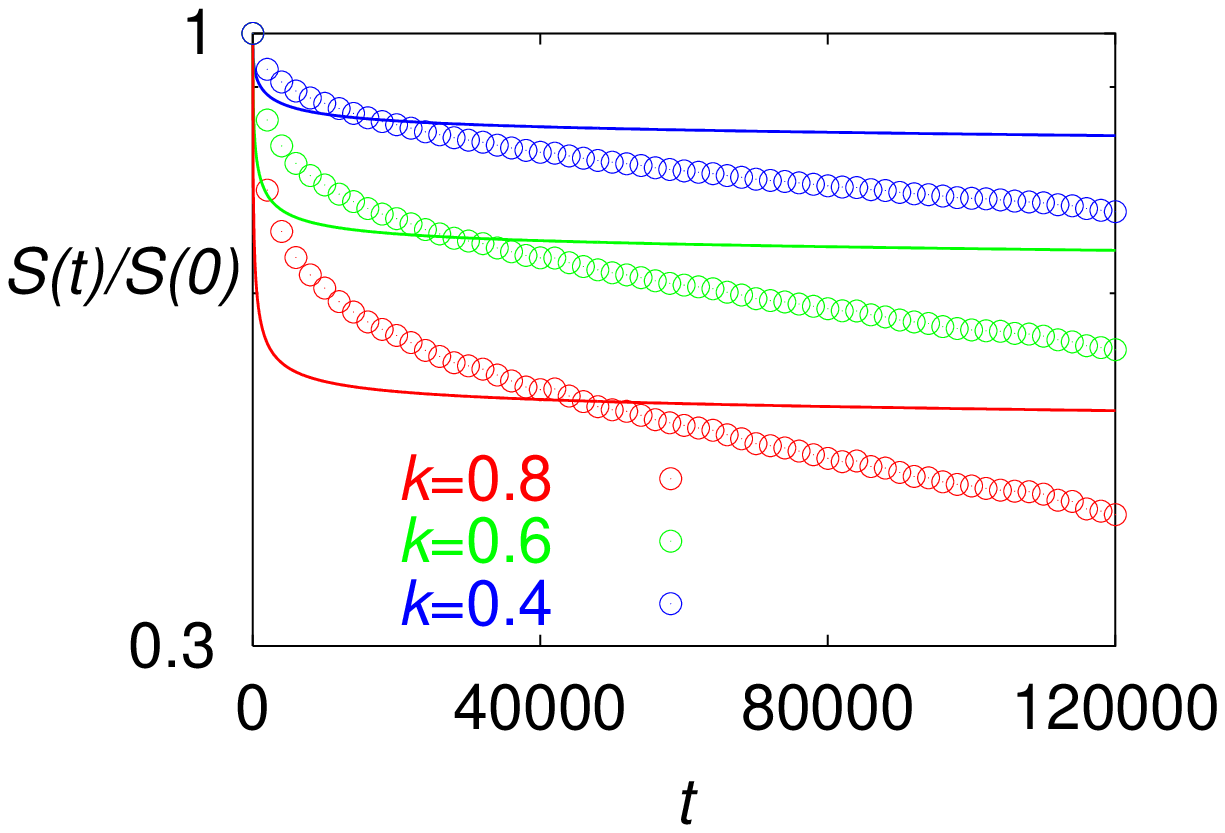}
  \caption{Normalized dynamic structure factors for $N=200$ and $k=0.4,\,0.6$,
    and 0.8. for a) $l_p=1.5$ and b) $l_p=5$. Note the different scales on the
    axes.} 
  \label{fig:dynstru}
\end{figure}
In spite of this uncertainty the difference is large enough to conclude that
also in this case the length scale governing the motion pattern, here $d_T$,
decreases strongly with increasing stiffness. 

Our simulations show that the lengths commonly associated with a change in
chain diffusion mechanism, the entanglement length $N_e$ and the tube diameter
$d_T$, decrease as the intrinsic stiffness of the chains increases. As a
result, the motion patterns change qualitatively already for small stiffness
(persistence length $l_p=1.5-3$ monomer diameters) and become completely
different from that of flexible chains when $N_e$ and/or $d_T$ become
comparable to or even smaller than $l_p$ (as is the case for $l_p=5$). For
$l_p=5$, the cooperation of entanglements and stiffness obstructs isotropic
random walk motion on all scales, the Rouse regime disappears and the chain
reptates along its narrow tube from the beginning. The absence of Rouse
dynamics distinguishes this situation from the standard de~Gennes - Doi -
Edwards picture~\cite{degennes71,doi86}. We suggest the term {\it strong
reptation} for this scenario. A similar transition has been proposed by
Morse~\cite{morse98a} who distinguishes between {\it loose} and {\it tight}
entanglement. One should note that most real polymers do have persistence
lengths larger than one monomer diameter. Even something commonly considered
flexible like polyisoprene has $l_p=1.5-2$~\cite{faller00a,faller00sb}. Our
results suggest that even to such polymers the standard Rouse/reptation model
can only be applied with caution. We have found deviations from the
predictions of the standard model also for the reorientation of local chain
segments~\cite{faller00b}, which are in good agreement with NMR measurements
on polybutadiene melts~\cite{graf98}.

None of the existing theoretical models can completely explain the observed
behavior. In the low persistence length region the packing
length concept~\cite{fetters94} gives a useful power law dependence
$N_e\propto l_p^{-2}$ but in the region where the two lengths become comparable
there is not yet an analytic understanding.  
\section*{Acknowledgements}
We thank B. D{\"u}nweg, R. Everaers, A. Heuer, K. Kremer, 
and M. P{\"u}tz for fruitful discussions. Financial support from the German
ministry of research (BMBF) is gratefully acknowledged.

\end{document}